\documentclass[twoside,english,sort&compress]{iopart}
\PassOptionsToPackage{natbib=true}{biblatex}
\usepackage[T1]{fontenc}
\usepackage[latin9]{inputenc}
\usepackage{geometry}
\usepackage{xcolor}
\usepackage{pdfcolmk}
\usepackage{babel}
\usepackage{array}
\usepackage{booktabs}
\usepackage{multirow}
\usepackage{amsbsy}
\usepackage{amstext}
\usepackage{graphicx}
\PassOptionsToPackage{normalem}{ulem}
\usepackage{ulem}
\usepackage[unicode=true,
 bookmarks=true,bookmarksnumbered=false,bookmarksopen=false,
 breaklinks=false,pdfborder={0 0 1},backref=false,colorlinks=true]
 {hyperref}
\hypersetup{
 pdfauthor={Mr. Wan},
  citecolor=blue,linkcolor=blue,urlcolor=blue}

\makeatletter

\providecommand{\tabularnewline}{\\}

\providecolor{lyxadded}{rgb}{0,0,1}
\providecolor{lyxdeleted}{rgb}{1,0,0}
\DeclareRobustCommand{\lyxadded}[3]{{\texorpdfstring{\color{lyxadded}{}}{}#3}}
\DeclareRobustCommand{\lyxdeleted}[3]{{\texorpdfstring{\color{lyxdeleted}\lyxsout{#3}}{}}}
\DeclareRobustCommand{\lyxsout}[1]{\ifx\\#1\else\sout{#1}\fi}

\usepackage{iopams}
\usepackage{setstack}

\newcommand{\eqref}[1]{(\ref{#1})}
\usepackage{fancyhdr}
\usepackage[justification=raggedright,labelfont=bf,singlelinecheck=true]{caption}
\usepackage{tablefootnote}

\makeatother

\usepackage[numbers]{natbib}
\begin{document}
\title{A high-fidelity surrogate model for the ion temperature gradient (ITG)
instability using a small expensive simulation dataset}
\author{Chenguang Wan$^{1,4}$, Youngwoo Cho$^{1}$, Zhisong Qu$^{1,*}$,
Yann Camenen$^{3}$, Robin Varennes$^{1}$, Kyungtak Lim$^{1}$, Kunpeng
Li$^{1}$, Jiangang Li$^{4}$, Yanlong Li$^{4}$, and Xavier Garbet$^{1,2,*}$}
\address{1. School of Physical and Mathematical Sciences, Nanyang Technological
University, Singapore 637371, Singapore}
\address{2. CEA, IRFM, F-13108 Saint Paul-lez-Durance, France}
\address{3. Aix-Marseille Universit�, CNRS, PIIM UMR 7345, Marseille, France}
\address{4. Institute of Plasma Physics, Hefei Institutes of Physical Science,
Chinese Academy of Sciences, Hefei 230031, China}
\ead{\href{mailto:zhisong.qu@ntu.edu.sg}{zhisong.qu@ntu.edu.sg} and \href{mailto:xavier.garbet@ntu.edu.sg}{xavier.garbet@ntu.edu.sg}}
\begin{abstract}
One of the main challenges in building high-fidelity surrogate models
of tokamak turbulence is the substantial demand for high-quality data.
Typically, producing high-quality data involves simulating complex
physical processes, which requires extensive computing resources.
In this work, we propose a fine tuning-based approach to develop the
surrogate model that reduces the amount of high-quality data required
by 80\%. We demonstrate the effectiveness of this approach by constructing
a proof-of-principle ITG surrogate model using datasets generated
from two gyrokinetic codes, GKW and GX. GX needs in terms of computing
resources are much lighter than GKW. Remarkably, the surrogate models'
performance remain nearly the same whether trained on 798 GKW results
alone or 159 GKW results plus an additional 11979 GX results. These
encouraging outcomes indicate that fine tuning methods can significantly
decrease the high-quality data needed to develop the simulation-driven
surrogate model. Moreover, the approach presented here has the potential
to facilitate surrogate model development for heavy codes and may
ultimately pave the way for digital twin systems of tokamaks.
\end{abstract}
\submitto{\NF }
\maketitle

\section{Introduction}

Predicting the temperature and density of the core plasma in tokamaks
is crucial for optimizing discharge scenarios, interpreting fusion
experiments, and designing future fusion devices. Typically, time-evolved
simulations of tokamak discharges over the course of a pulse are performed
using an \textquotedbl integrated modeling\textquotedbl{} approach
, such as ETS \citep{Falchetto2014}, PTRANSP \citep{Budny2008},
TSC \citep{Kessel2006}, TRANSP \citep{Breslau2018}, CRONOS \citep{Artaud2010},
JINTRAC \citep{Romanelli2014}, METIS \citep{Artaud2018}, ASTRA \citep{Pereverzew1991},
TOPICS \citep{Hayashi2010}, etc. These approaches integrate multiple
module codes, each addressing a different physical process in tokamak
plasmas, into a single workflow. A key component of these integrated
models is the prediction of turbulent fluxes, particularly in the
core of the tokamak, where plasma microinstabilities often dominate
transport. However, calculating these fluxes using nonlinear gyrokinetic
models is computationally demanding for routine simulation of tokamak
discharge evolution.

The plasma physics community has developed numerous first-principles
or empirical physics codes to address these challenges under varying
fidelity conditions. Gradient-driven codes include GENE \citep{Jenko2000},
GKW \citep{Peeters2009}, GS2 \citep{Kotschenreuther1995}, GX \citep{Mandell2022},
and GYRO \citep{Candy2003}. In addition, there are flux-driven codes
such as GYSELA \citep{Grandgirard2006} and GT5D \citep{Idomura2005}.
Some physical-point-acceleration codes include QuaLiKiz \citep{Citrin2017,Bourdelle2016}
and TGLF \citep{Staebler2007}. These simulation codes are effective,
but there is always a trade-off between physical fidelity and computational
efficiency. Basically, high physical fidelity often makes simulations
computationally expensive for routine tokamak simulations and vice
versa. However, scenarios such as uncertainty quantification, scenario
optimization, and controller design demand fast, high-fidelity simulation
codes, particularly for many-query applications like multi-dimensional
integrated models (e.g., \citep{Poli2018,Felici2011,Artaud2018,Romanelli2014}),
where both speed and accuracy are paramount.

Recently, the machine learning (ML) method has emerged as a powerful
alternative for new model development by learning from simulation
or experimental datasets, often referred to as surrogate models. ML
surrogate models can capture the complex relationships between plasma
parameters and tokamak response. To date, many surrogate models based
on neural networks have been developed to solve different tasks in
tokamaks, such as tokamak operation \citep{Abbate2024}, integrated
modeling \citep{Boyer2019} etc. Based on the data type usage, these
surrogate,models can be divided into two categories: experimental
data-driven and simulation data-driven models.

Experimental data-driven surrogate models have achieved significant
advancements, including disruption prediction \citep{Zheng2023,Kates-Harbeck2019},
tokamak operation based on reinforcement learning \citep{Murari2024,Seo2024},
last closed-flux surface reconstruction \citep{Wan2023,Wan2024},
entire discharge estimation \citep{Wan2021,Wan2022}. These methods
are accurate and capable of reconstructing the discharge process of
an operating tokamak based on input parameters. However, experimental
data-driven models face challenges in interpreting experimental results
due to the lack of theoretical models. Furthermore, it seems impractical
to directly apply them to the design or operation of \emph{new} devices.

Simulation data-driven surrogate models run efficiently and have the
capability to explain physical processes in tokamak experiments, as
well as aid in designing \emph{new} tokamak devices. Due to these
advantages, simulation data-driven surrogate models have been receiving
increasing attention. Several efforts have been made in this area,
including the quasilinear turbulent transport codes, TGLF ($\sim2$
CPUs per case) \citep{Meneghini2017,Staebler2007,VanDePlassche2020}
and QualiKiz ($\sim30$ CPUs per case) \citep{Citrin2017,VanDePlassche2020},
the pedestal confinement code EPED ($\sim10$ CPUh per case) \citep{Meneghini2017,Staebler2007},
the neutral beam heating code NUBEAM ($\sim1$ CPUh per case) \citep{Goldston1981,Pankin2004,Boyer2019},
and the tokamak edge plasma code SOLPS, which can take months of wall-clock
time for high-fidelity simulations \citep{Bonnin2016,Dasbach2023}.
Successful simulation-driven surrogate model typically fall into two
categories: (1) fast-running models like TGLF, QualiKiz, EPED, and
NUBEAM, which can efficiently generate large datasets, or (2) models
such as SOLPS-ITER, where extensive datasets are already available.
However, developing high-fidelity nonlinear simulation-driven surrogate
models from scratch presents significant challenges due to the prohibitive
computational cost of generating large training datasets. For instance,
a simple ITG case using GKW requires $\sim300$ CPU hours, while developing
a typical 4D surrogate model like QLKNN-4D \citep{Citrin2015} generally
needs $\sim10^{6}$ cases. Acquiring such a large amount of high-fidelity
data is often either infeasible or highly resource-intensive.

The computational cost of nonlinear gyrokinetic simulation codes is
typically high, making it an urgent challenge to build a surrogate
model with a small dataset. Developing surrogate models for gyrokinetic
simulation codes using a sizable, high-fidelity nonlinear dataset
is practically infeasible. Several innovative approaches have been
explored to address this issue. These include adaptive sampling training
database \citep{Zanisi2024}, hybrid machine learning and physical
simulations to reduce the data requirements \citep{Citrin2024}, physics-constrained
neural networks \citep{Arnaud2024}, Generative Artificial Intelligence
(GAI) \citep{Clavier2024}, and neural partial differential equation
(PDE) surrogates \citep{Poels2023}. These methods incorporate physical
prior knowledge to varying degrees, effectively reducing the data
requirements for high-fidelity surrogate models. However, to the best
of the authors' knowledge, few studies \citep{Fuhr2023} have explored
leveraging large amounts of low-fidelity data to enhance the performance
of surrogate models on the high-fidelity data.

In this work, we aim to address this problem by introducing a novel
surrogate model-building approach as a proof of concept. We begin
by generating a sizable dataset of low-fidelity, rapid simulations
using fast codes with low resolution, followed by a smaller dataset
of high-fidelity, high-resolution simulations to refine the model.
This strategy strikes a balance between precision and computational
efficiency, enabling the development of effective gyrokinetic surrogate
models. Specifically, we are focusing on constructing a surrogate
model for ion thermal diffusivity $\chi_{i}$ and wave number spectrum
of ITG modes using GKW \citep{Peeters2009} and GX \citep{Mandell2022}.

All simulations are run with 1 ion species (D) and adiabatic electrons
with scanning four parameter inputs: safety factor $q$, magnetic
shear $\hat{s}$, ion temperature gradient $R/L_{T_{i}}$, and density
gradient $R/L_{n}$. The outputs are ion thermal diffusivity $\chi_{i}$
and wave number spectrum. We collected 13310 simulation results using
the GX code to primarily train our machine learning (ML) model and
generated 1267 high-fidelity results using the GKW code to fine-tune
and validate the model's accuracy.

The surrogate model, developed with only 159 GKW results and 11979
GX results, achieved the same performance as a model trained with
798 GKW results alone. This approach reduced the requirement for high-resolution,
computationally expensive results by 80\% while maintaining equivalent
accuracy. Furthermore, the surrogate model, based on a neural network,
is orders of magnitude faster than direct simulations, requiring only
approximately one milliseconds per case, making it suitable for real-time
modeling.

The present paper is structured as follows. Section \ref{sec:Methods}
details the dataset preparation, the methodology for constructing
the machine learning model, and the training procedures. Section \ref{sec:Results}
highlights the performance of the surrogate model, including a comparative
analysis with other models. Finally, Section \ref{sec:Discussion}
provides a brief discussion and conclusion.

\section{Methods \label{sec:Methods}}

In this section, we provide detailed explanations of the dataset,
machine learning model construction, and model training process.

\subsection{Dataset generation}

To generate a database for ITG instabilitiy, nonlinear gyrokinetic
simulations are performed using both GKW and GX codes. GKW \citep{Peeters2009}
is a CPU-based code that solves turbulent transport problems on a
fixed grid in five-dimensional space, employing a combination of finite
difference and pseudo-spectral methods. It utilizes Fourier pseudo-spectral
techniques in spatial dimensions and is a high-fidelity and computationally
intensive code. In our ITG instability simulations, each GKW case
required approximately 300 CPU hours. GX \citep{Mandell2022} is a
GPU-native code designed for solving the nonlinear gyrokinetic system
governing low-frequency turbulence in magnetized plasmas, including
those in tokamaks and stellarators. It employs a Fourier-Laguerre-Hermite
pseudo-spectral formulation in phase space, offering a fast gyrokinetic
solver that requires only $\sim0.25$ GPU hours per case for our ITG
instability simulations. While GX and GKW provide comparable fidelity,
we adjusted the resolution in velocity space for GX and slightly modified
other parameters that lower the fidelity of the resulting simulation
compared with our GKW cases. (see table \ref{tab:GKW_GX_settings}).
All other settings were kept consistent with those used in GKW, such
as flux surface geometry of Miller equilibrium, elongation of flux
surface equal to 1, collision frequency to 0, etc. Notably, when GX
was configured to closely match GKW, our test case indicated that
GX required $\sim12$ GPU hours per case.

\begin{table}
\caption{The difference between GKW and GX in our simulation. \label{tab:GKW_GX_settings}}

\centering{}%
\begin{tabular}{>{\raggedright}p{0.25\linewidth}>{\raggedright}p{0.2\linewidth}>{\raggedright}p{0.3\textwidth}}
\toprule 
\multirow{1}{0.25\linewidth}{} & GKW & \multirow{1}{0.3\textwidth}{GX}\tabularnewline
\midrule
Architecture & CPU-based & GPU-based\tabularnewline
Spectral method in space & Fourier Pseudo-spectral in space & \multirow{1}{0.3\textwidth}{Fourier Laguerre Hermite pseudo spectral formulation in phase space}\tabularnewline
Speed per case & \textbf{$\sim\boldsymbol{300}$} CPU hours & $\sim\boldsymbol{0.25}$ GPU hours\tabularnewline
Parallel $V_{\parallel}$ mesh points & 64 & 8\tabularnewline
$\mu$ mesh points & 8 & 4\tabularnewline
Max. time steps & 1200 & 400\tabularnewline
\bottomrule
\end{tabular}
\end{table}

The input space of GX and GKW codes ( $\sim15$ dimensions for typical
simulations) is too large with brute-force scans, especially for a
demonstrative proof-of-principle study. Therefore, we reduced the
input parameter space to a subset that significantly impacts our goal:
the ITG mode. The selected input parameters and their ranges, shown
in table \ref{tab:4D_input_space}, include the ion temperature gradient
($R/L_{T_{i}}$), density gradient ($R/L_{n}$), safety factor ($q$),
magnetic shear ($\hat{s}$).

The dataset comprises 1267 GKW and 13310 GX simulations. The simulations
required approximately 0.39 million CPU hours (MCPUh) and 0.33 million
GPU hours (MGPUh). The input space was structured as a rectangular,
uniform four-dimensional grid, with bounds covering the parameter
regimes typically encountered in the core of standard aspect-ratio
present-day tokamaks and future devices such as ITER and DEMO. The
dataset, stored in HDF5 format, occupies approximately 4 GB of storage.

\begin{table}
\begin{centering}
\caption{4D hyperrectangle bounds and number of points of the GX and GKW dataset.
Each input parameter is uniformly distributed to widely verify our
methodology. \label{tab:4D_input_space}}
\par\end{centering}
\centering{}%
\begin{tabular}{ccccl}
\toprule 
Variable & GKW points & GX points & Min & Max\tabularnewline
\midrule
Ion temperature gradient, $R/L_{T_{i}}$ & 6 & 11 & 4.5 & 7\tabularnewline
Density gradient,$R/L_{n}$ & 6 & 11 & 0.5 & 3.0\tabularnewline
Safety factor, $q$ & 6 & 11 & 1 & 3\tabularnewline
Magnetic shear, $\hat{s}$ & 6 & 10 & 0.1 & 1\tabularnewline
GKW valid calculations & 1267 & - & \multicolumn{2}{c}{$\approx$ 0.39 MCPUh}\tabularnewline
GX valid calculations & - & 13310 & \multicolumn{2}{c}{$\approx$ 0.33 kGPUh}\tabularnewline
\bottomrule
\end{tabular}
\end{table}

\subsection{Data preprocessing}

Once the data is generated, we preprocess the datasets to extract
key information and restructure them to ensure compatibility with
the machine learning models. The data preprocessing workflow can be
divided into three steps: data cleaning, reduction, and splitting.

\subsubsection{Data cleaning}

Gyrokinetic nonlinear simulations are inherently complex and numerically
challenging, requiring intricate setups and often encountering numerical
instabilities introduced by nonlinearities in the model. Some simulations
may fail or not converge. In this paper, we apply three simple criteria
to determine the validity of our simulations:
\begin{enumerate}
\item The total simulation steps exceeds 600 for GKW, and 200 for GX.
\item The simulation reaches nonlinear saturation beginning at or before
60\% of the total simulation duration.
\item The $k_{x}$ spectrum $\phi_{x}$ and $k_{y}$ spectrum $\phi_{y}$
have over 20 points and 10 points in the ranges of $\ensuremath{[-1.6,1.6]}$
and $[0,0.75]$, respectively. These ranges are determined by the
wavenumbers that researchers are interested in and the significant
linear growth rate spans \citep{Dimits2000}.
\end{enumerate}
A simulation is considered invalid if it terminates too early, as
this typically indicates numerical instability or computational failure.
Additionally, if the nonlinear saturation phase is too short, averaging
quantities $\chi_{i}$, $\phi_{x}$, $\phi_{y}$ will be inaccurate,
over the final 40\% of the simulation length may lead to inaccurate
results. Finally, if the number of points in the wavenumber intervals
predicted by the machine learning model is insufficient, the surrogate
model\textquoteright s accuracy and reliability will be significantly
affected. Notably, the intervals {[}-1.6, 1.6{]} of $k_{x}$, and
{[}0, 0.75{]} of $k_{y}$ is a choice to balance the computational
resource and physical requirements. If the $k_{x}$, and $k_{y}$
regime is excessively broad, the computational resource requirements
increase significantly. Because in order to simplify the surrogate
model, we fixed the dimensionality of the output. The representative
case in this context is the cyclone case. In this case, for the toroidal
mode number relevant to ITG, $k_{y}$ corresponds to approximately
0.5. As shown in \citep{Dimits2000}, the region exhibiting a significant
linear growth rate spans from 0 to 0.7. Hence, we set the range of
$k_{y}$ from 0 to 0.75.

\subsubsection{Data reduction}

The raw simulation data from GKW and GX consists of time sequences,
with sequence lengths depending on the simulation time steps. In this
study of ITG simulations, we focus only on the final effects of the
wave spectrum and thermal diffusivity. First, we calculate the average
values of $\chi_{i}$, $\phi_{x}$, and $\phi_{y}$ over the last
40\% of the simulation time steps. Next, we apply linear interpolation
to $\phi_{x}$ and $\phi_{y}$ to align these values. After interpolation,
$\phi_{x}$ and $\phi_{y}$ contain 101 and 51 points, respectively,
covering $k_{x}$ and $k_{y}$ the ranges of {[}-1.6, 1.6{]} and {[}0,
0.75{]}. Finally, we take the logarithmic scale of $\phi_{x}$, and
$\phi_{y}$ . With these processing steps, the target of our model
will have 153 channels: 1 channel for $\chi_{i}$, 101 channels for
$\phi_{x}$, and 51 channels for $\phi_{y}$.

\subsubsection{Data splitting}

In this paper, we use two code GKW and GX simulation results. Firstly,
we randomly split the GKW dataset into three subsets: $S_{1}$ , $S_{2}$,
and $S_{3}$, containing 63\%, 7\%, and 30\% of the data, which correspond
to 798, 89, and 380 GKW simulation results, respectively. Next, we
randomly divide the GX simulation dataset into $S_{4}$ and $S_{5}$
with proportions of 90\% and 10\%, encompassing 11979 and 1331 GX
simulation results, respectively. Notably, $S_{3}$ is a slightly
larger test set proportion compared to those typically used in machine
learning studies, as we aim to evaluate our method across a broader
parameter space using a extensive dataset.

\subsection{Machine learning model \label{subsec:ML_Model}}

\begin{figure}
\begin{centering}
\includegraphics[width=0.75\textwidth]{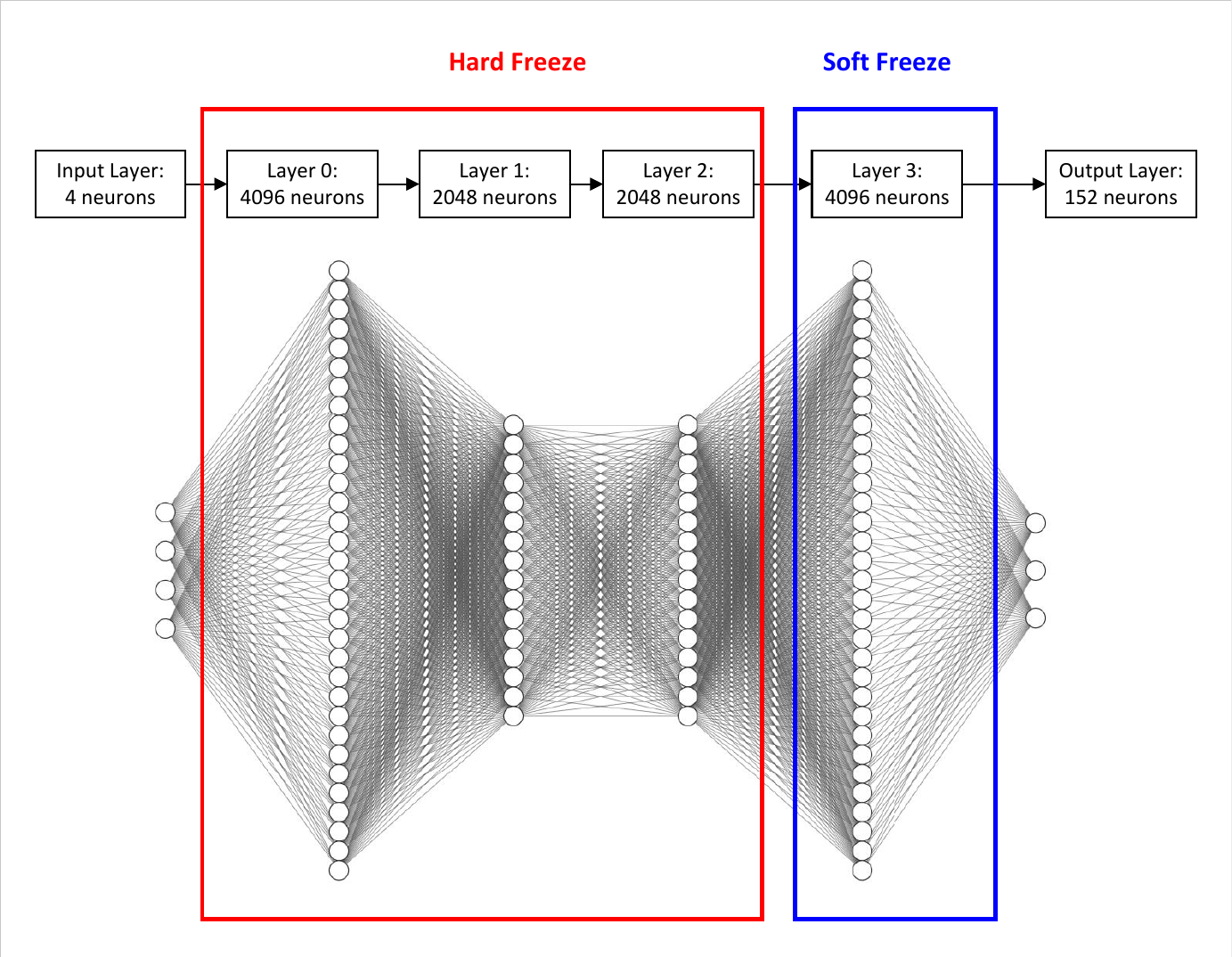}
\par\end{centering}
\caption{Fine-tuning technology in our MLP model. The red box is defined with
a zero learning rate, while the blux box, labeled \textquotedbl Soft
freeze\textquotedbl , is set to a learning rate of $1\times10^{-5}$.
\textbf{\label{fig:FineTune}}}
\end{figure}

Recently, the remarkable success of generative artificial intelligence
applications like Chat-GPT \citep{Brown2020,Radford2019} and Stable
Diffusion \citep{Rombach2022} has highlighted the effectiveness of
fine-tuning. Fine-tuning involves leveraging knowledge from a large,
low-accuracy dataset and refining it with a smaller, high-accuracy
dataset to enhance model performance for a specific task. As illustrated
in figure \ref{fig:FineTune}, for the fine-tuned model, we apply
a similar fine-tuning approach to train a gyrokinetic surrogate model,
utilizing fast, low-resolution simulation data to improve the model\textquoteright s
performance on high-resolution, low-speed simulations. Specifically,
for fine-tuning, we follow these steps:
\begin{enumerate}
\item \textbf{Initial Training: }Train the entire neural network using $S_{4}$
and $S_{5}$ as the training and validation sets with a learning rate
of $1\times10^{-3}$
\item \textbf{Model Selection: }Select the best-performing model based on
validation set performance.
\item \textbf{Layer Freezing: }Apply a \textquotedblleft hard freeze\textquotedblright{}
to the first three layers (setting their learning rate to 0) and a
\textquotedblleft soft freeze\textquotedblright{} to the last layer
(setting its learning rate to $5\times10^{-5}$).
\item \textbf{Re-training:} Re-train the neural network using 20\% $S_{1}$
and 20\% $S_{2}$ as training and validation sets.
\item \textbf{Testing: }Test the model on set $S_{3}$.
\end{enumerate}
For the GX-only and the GKW-only simulation result training models,
we do something similar, but only use steps 1, 2, and 5. And in the
GKW simulation trained model, in step 1 we use $S_{1}$, $S_{2}$
instead of $S_{4}$, $S_{5}$.

In our approach, the \textquotedblleft hard freeze\textquotedblright{}
(learning rate of 0) prevents specific layers from being modified
by the model trained with the new data, while the \textquotedblleft soft
freeze\textquotedblright{} (learning rate of $1\times10^{-5}$) allows
limited adjustments to align the model with new data. In our case,
the new data refers to GKW simulation results. This strategy assumes
that the earlier layers capture general lower-level knowledge suitable
for gyrokinetic ITG simulations, whereas the later layers are more
likely to discern high-level knowledge relevant to the specific task
for which the model is being trained.

\begin{table}
\caption{The loss comparison among multiple models. The MLP is the simplest
neural network model. The CNN is a convolution-based model commonly
used in image processing. Transformer and Transformer Encoder are
attention-based models widely used in large language model applications.
Based on the MSE loss comparison, the MLP model achieved the best
performance. \label{tab:loss-compare}}

\centering{}%
\begin{tabular}{cc}
\toprule 
Model type & Mean Squared Error (MSE) loss\tabularnewline
\midrule
Multilayer perceptron (MLP) & $\sim1\ensuremath{\times}10^{-4}$\tabularnewline
Transformer & $\sim1\ensuremath{\times}10^{-2}$\tabularnewline
Transformer Encoder & $\sim1\ensuremath{\times}10^{-2}$\tabularnewline
Convolutional Neural Networks (CNN) & $\sim1\ensuremath{\times}10^{-3}$\tabularnewline
\bottomrule
\end{tabular}
\end{table}

In this paper, we compare the training results of the MLP model with
three other popular models: CNN, Transformer, and Transformer Encoder
on GX data, all of which have the same parameter size of $\sim10$
million. MLP (Multilayer Perceptron): This is the simplest type of
neural network, consisting of fully connected layers. Its advantages
include ease of implementation and interpretability, making it suitable
for problems with structured numerical data. However, it may struggle
to capture spatial structures or long-range dependencies. CNN (Convolutional
Neural Network): This model leverages convolution operations and is
commonly used in image processing tasks. It is effective at capturing
local spatial features and has translation invariance, making it useful
when data has spatial or local correlations. However, it can require
careful design to handle non-spatial or sequential data efficiently.
Transformer and Transformer Encoder: These are attention-based models
widely used in large language models and sequence modeling tasks.
Their main advantage is their ability to capture long-range dependencies
and context within data effectively. However, they typically require
substantial data and computational resources, and can be more complex
to train compared to simpler architectures. Table \ref{tab:loss-compare}
shows that MLP has the lowest MSE loss value. Generally, in machine
learning model evaluation, the smaller the loss value, the better
performance of the model. Therefore, we choose MLP for the present
work. This choice is also intuitive, as this study serves as a proof-of-principle
with only four input parameters (4 channels) and three output parameters
(153 channels, with $\chi_{i}$, $\phi_{x}$, and $\phi_{y}$ having
1, 101, and 51 channels, respectively), making their mapping relationship
relatively straightforward. More complex, advanced models tend to
be harder to converge and typically require larger datasets for effective
training.

Our machine learning model was developed using PyTorch on Red Hat
Enterprise Linux 8, running on four A100 GPUs. During model training,
we utilized the Tree-structured Parzen Estimator (TPE) algorithm \citep{Bergstra2011}
to perform the architectural hyperparameter search. Additionally,
we experimented with various optimizers and regularization techniques,
ultimately identifying the optimal set of hyperparameters, as shown
in table \ref{tab:mr_hyperparameters}.

\begin{table}
\caption{\label{tab:mr_hyperparameters}Our model hyperparameters. Model architecture
can be found in figure \ref{fig:FineTune}}

\centering{}%
\begin{tabular}{ccc}
\toprule 
Hyperparameter & Explanation & Best value\tabularnewline
\midrule
$\text{\ensuremath{\eta}}_{t}$ & Learning rate in model train phase & $1\text{\ensuremath{\times10^{-3}}}$\tabularnewline
$\text{\ensuremath{\eta}}_{f}$ & Learning rate in model fine tuning phase & $5\times10^{-4}$\tabularnewline
$\text{Epoch}_{t}$ & Number of epochs in model train & 800\tabularnewline
$\text{Epoch}_{f}$ & Number of epochs in model fine tuning phase & 1000\tabularnewline
Optimizer & Optimizer type & Stochastic Gradient Descent (SGD)\tabularnewline
Loss & Loss function & Mean Squared Error (MSE) loss\tabularnewline
Scheduler & Scheduler type & OneCycle\citep{Smith2017}\tabularnewline
Dropout & Dropout probability & 0.1\tabularnewline
$B$ & Batch size & 4\tabularnewline
$C_{i}$ & Input channels & 4\tabularnewline
$C_{o}$ & Output channels & 152\tabularnewline
$E_{\text{mlp}}$ & Embed layers & {[}4096, 2048, 2048, 4096{]}\tabularnewline
\bottomrule
\end{tabular}
\end{table}

\section{Results \label{sec:Results}}

\begin{figure}
\begin{centering}
\begin{tabular}{cc}
\includegraphics[width=0.5\textwidth]{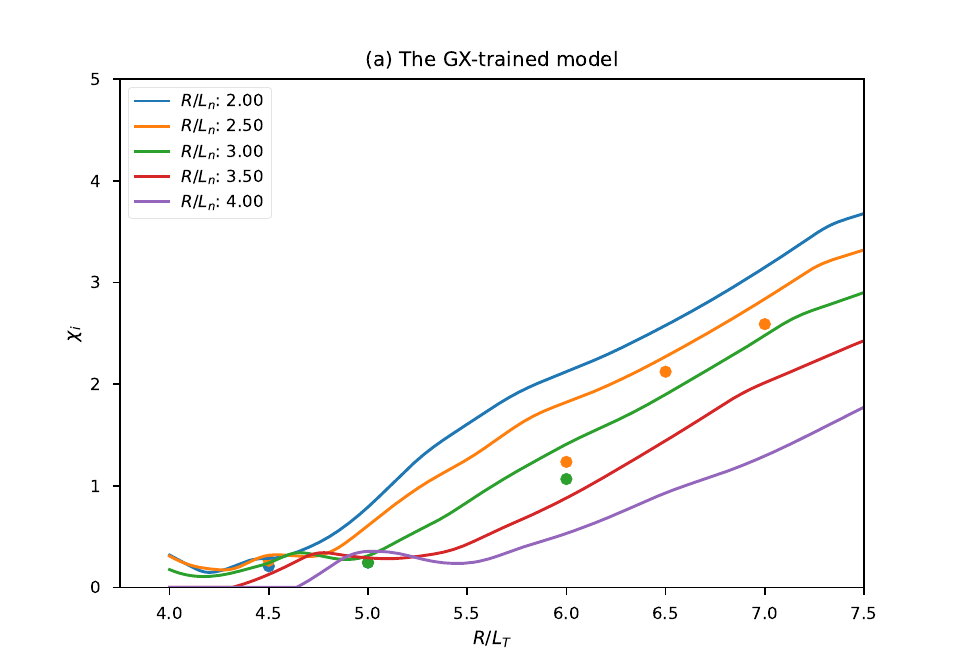} & \includegraphics[width=0.5\textwidth]{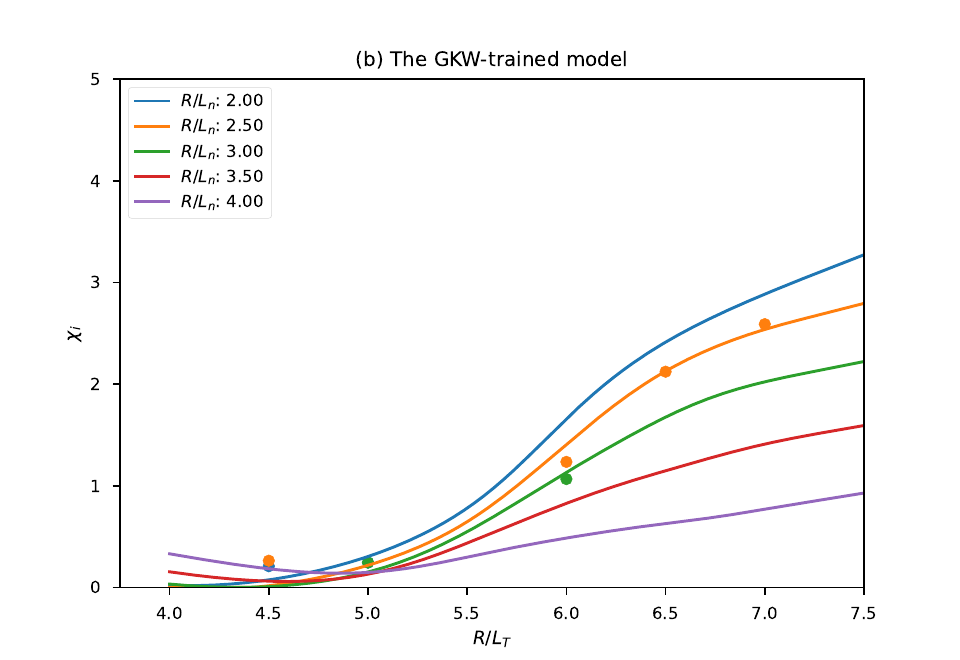}\tabularnewline
\includegraphics[width=0.5\textwidth]{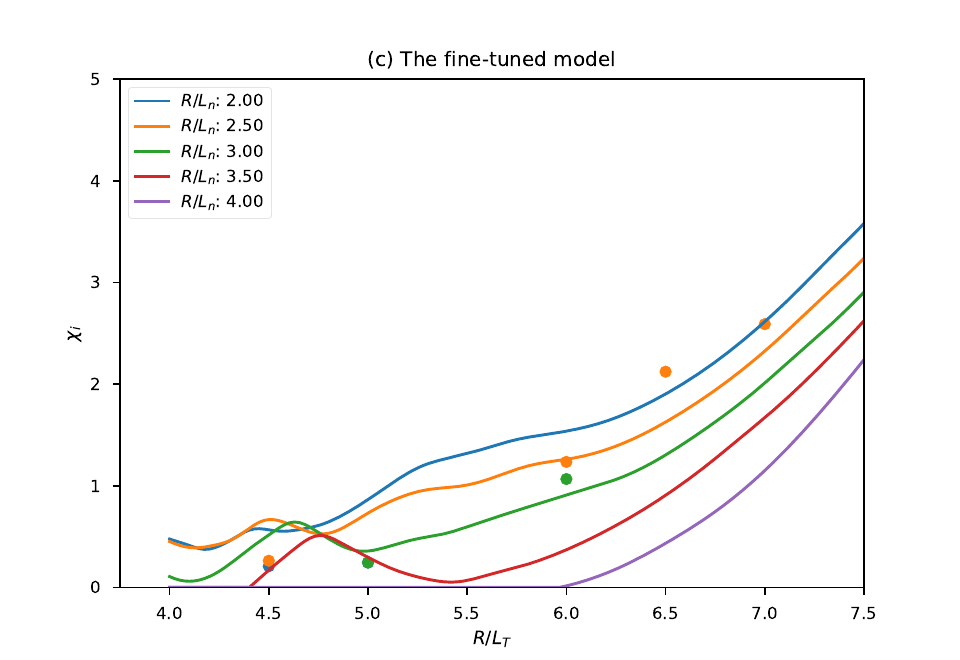} & \tabularnewline
\end{tabular}
\par\end{centering}
\centering{}\caption{\label{fig:diffvsrlt} The plots of $\chi_{i}$ versus $R/L_{T}$
with $q=1.4,$ $\hat{s}=0.64$. The solid lines $\chi_{i}$ vs. $R/L_{T}$
for different $R/L_{n}$. All circular points in three subplots indicate
actual GKW simulation results, with colors corresponding to the $R/L_{n}$
values of the solid lines. The behavior below the threshold (i.e.
non-zero flux below the threshold) will be problematic for any integrated
modeling application, making surrogate models inaccurate in this region.}
\end{figure}

After completing the model training, we obtained three models: the
GKW-trained model, the GX-trained model, and the fine-tuned model.
To evaluate the quality of these neural networks and assess the impact
of our underlying assumptions, we compared them with the actual GKW
simulation results for $\chi_{i}$ versus $R/L_{T}$ with $q=1.4$,
$\hat{s}=0.64$ in figure \ref{fig:diffvsrlt}. The figure illustrates
that the GX trained model does not align well with the real GKW labels,
whereas the GKW-trained model closely matches them. Despite employing
multiple techniques to prevent overfitting, the GKW trained model's
fit is almost too perfect, raising some concerns about potential overfitting.
In contrast, while the fine-tuned model is not as precise as the GKW-trained
model, it produces smoother curves in the $\chi_{i}$ vs. $R/L_{T}$
plot, particularly in regions far from the turbulence threshold. It
is important to note that the non-monotonic behavior of $\chi_{i}$
near the threshold may pose challenges for integrated modeling, as
abrupt variations in transport coefficients can compromise numerical
stability and convergence. Since this study serves as a proof of concept,
we prioritize clarity and ease of understanding in our methodology.
However, to mitigate this issue in future work, we will explore: (1)
Threshold labeling based on extensive simulation cases. (2) As shown
in \citep{Hornsby2024}, training a classifier to identify the threshold,
and (3) Restricting regression model training to the monotonic region
above the threshold. These improvements will enhance the monotonicity
of the surrogate model.

\begin{figure}
\begin{centering}
\includegraphics[width=1\textwidth]{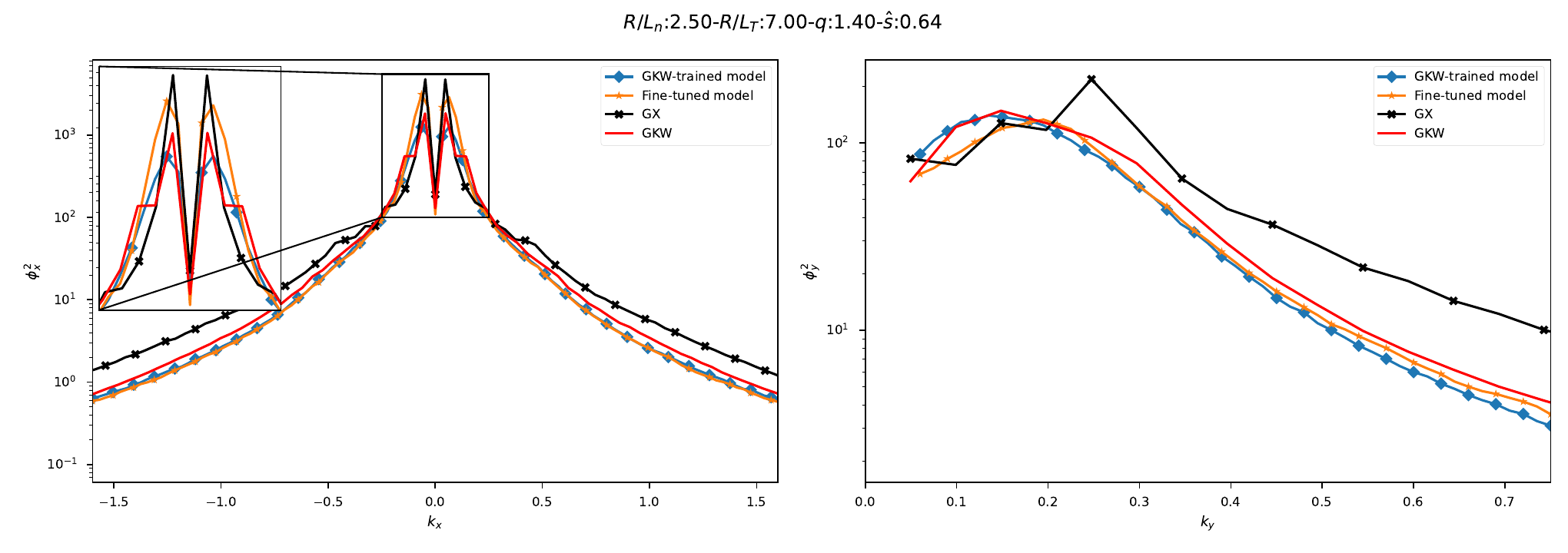}
\par\end{centering}
\caption{\label{fig:Pkxy}The $\phi_{x}$ and $\phi_{y}$ plots for GKW-trained
model, fine-tuned model, actual simulations of GKW and GX. The blue
diamonded solid line and the orange starred solid line represent the
predicted results of the GKW-trained and fine-tuned models, respectively.
The black crossed solid line and the red solid line correspond to
the actual simulations of GX and GKW, respectively.}
\end{figure}

Figure \ref{fig:Pkxy} presents a comparison of the two models, both
the GKW-trained and fine-tuned models, with typical GKW and corresponding
GX simulations, illustrating the $k_{x}$ spectrum $\phi_{x}$ and
the $k_{y}$ spectrum $\phi_{y}$. The fine-tuned and GKW-trained
models exhibit almost identical performance when the $k_{x}$ wavenumber
exceeds 0.25 or the $k_{y}$ wavenumber exceeds 0.1. In the zoom in
subplot of figure \ref{fig:Pkxy}, at low $k_{x}$ wavenumbers, the
fine-tuned model closely fits the GX simulation represented by the
black crossed solid line. This may explain why the fine-tuned model
overestimates $\phi_{x}$ in low $k_{x}$ wavenumbers. Overall, the
$\phi_{y}$ predictions from both surrogate models align well with
the actual GKW simulations across the entire range of $k_{y}$ wavenumbers.
In detail, the fine-tuned model performs slightly worse than the GKW-trained
model at low wavenumbers. The challenge arises because the GKW and
GX simulations do not have densely sampled data points at low wavenumbers.
We plan to address this issue in future studies by performing additional
simulations with denser data sampling in the low wavenumber range.

\begin{figure}
\begin{centering}
\includegraphics[width=0.8\textwidth]{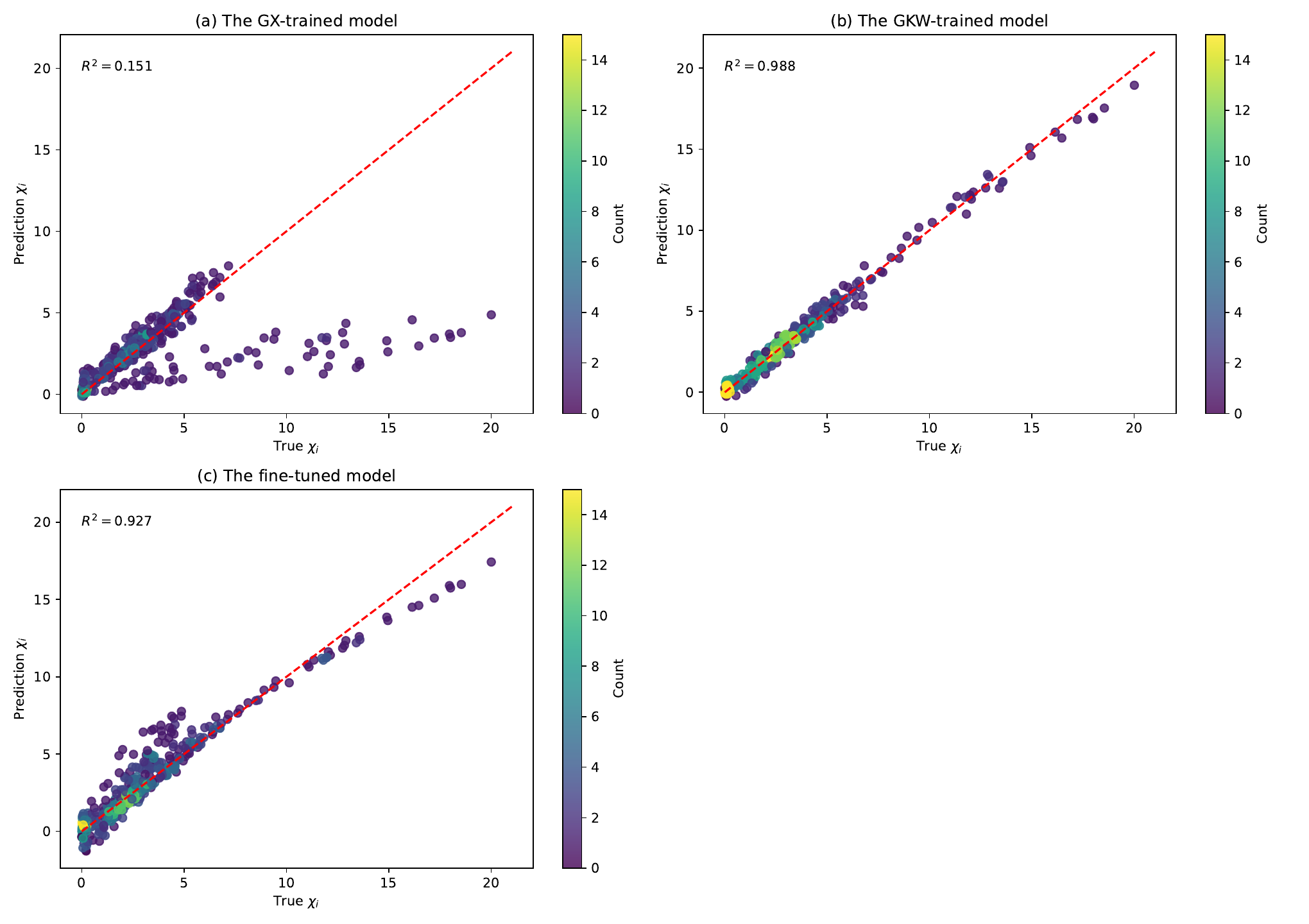}
\par\end{centering}
\centering{}\caption{\label{fig:Regression} Regression plots of $\chi_{i}$ for the three
models. The plots show that GX performs the worst, particularly when
$\chi_{i}$ is large. Fine-tuning significantly enhances GX\textquoteright s
performance, making points with small $\chi_{i}$ more concentrated
along the diagonal and those with large $\chi_{i}$ closer to it.}
\end{figure}

\begin{figure}
\begin{centering}
\begin{tabular}{cc}
\includegraphics[width=0.5\textwidth]{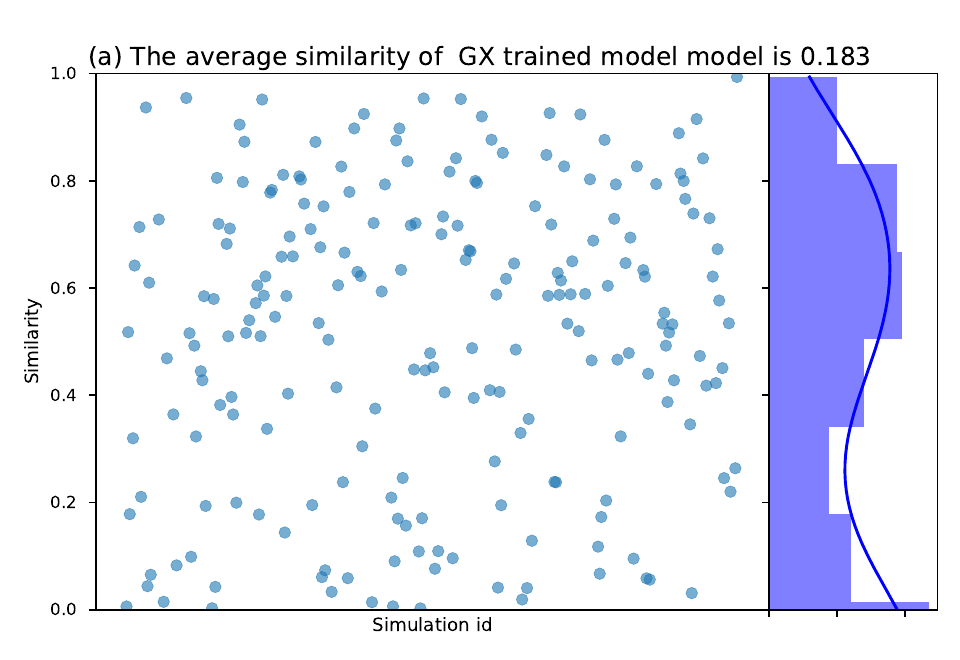} & \includegraphics[width=0.5\textwidth]{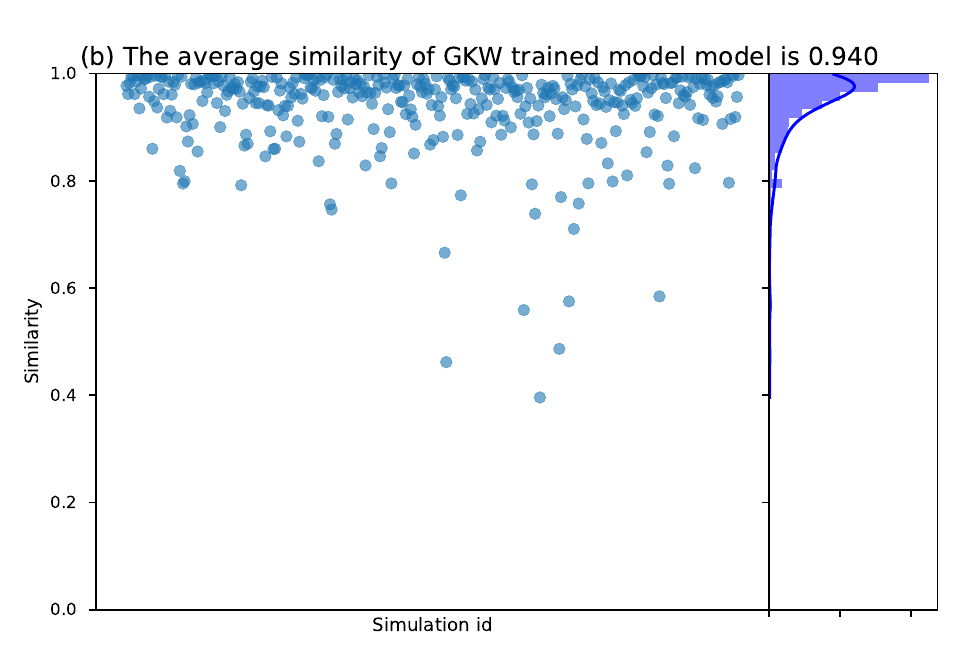}\tabularnewline
\includegraphics[width=0.5\textwidth]{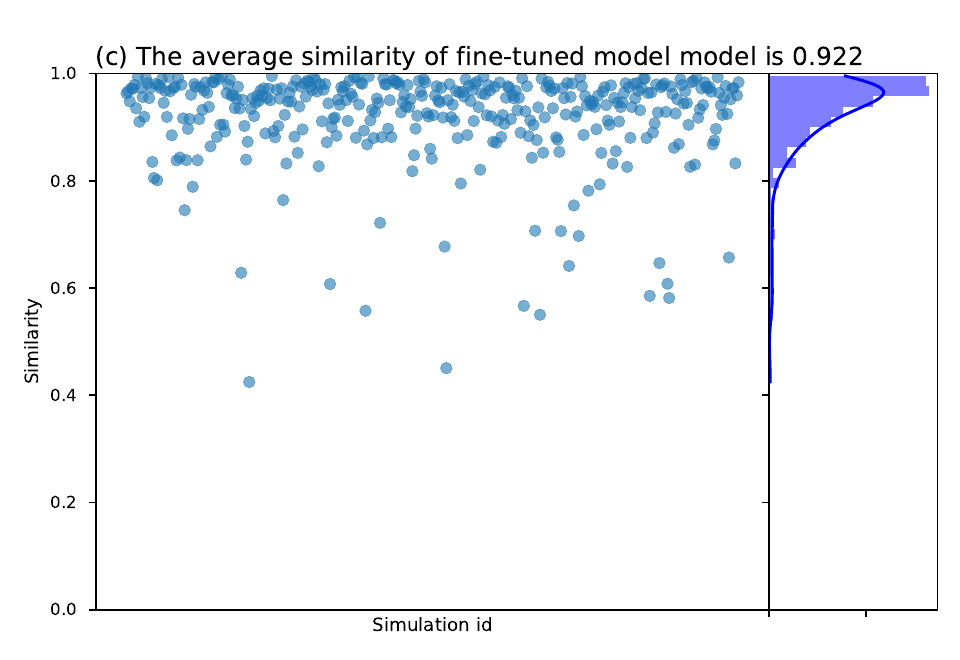} & \tabularnewline
\end{tabular}
\par\end{centering}
\centering{}\caption{\label{fig:simDis}The similarity distribution in the whole test set
of (a) GX-trained, (b) GKW-trained, (c) fine-tuned model predictions.
The similarity defined as equation \ref{eq:1}. To the right of each
subfigure are the histograms with density estimates.}
\end{figure}

\begin{figure}
\centering{}\includegraphics[width=1\textwidth]{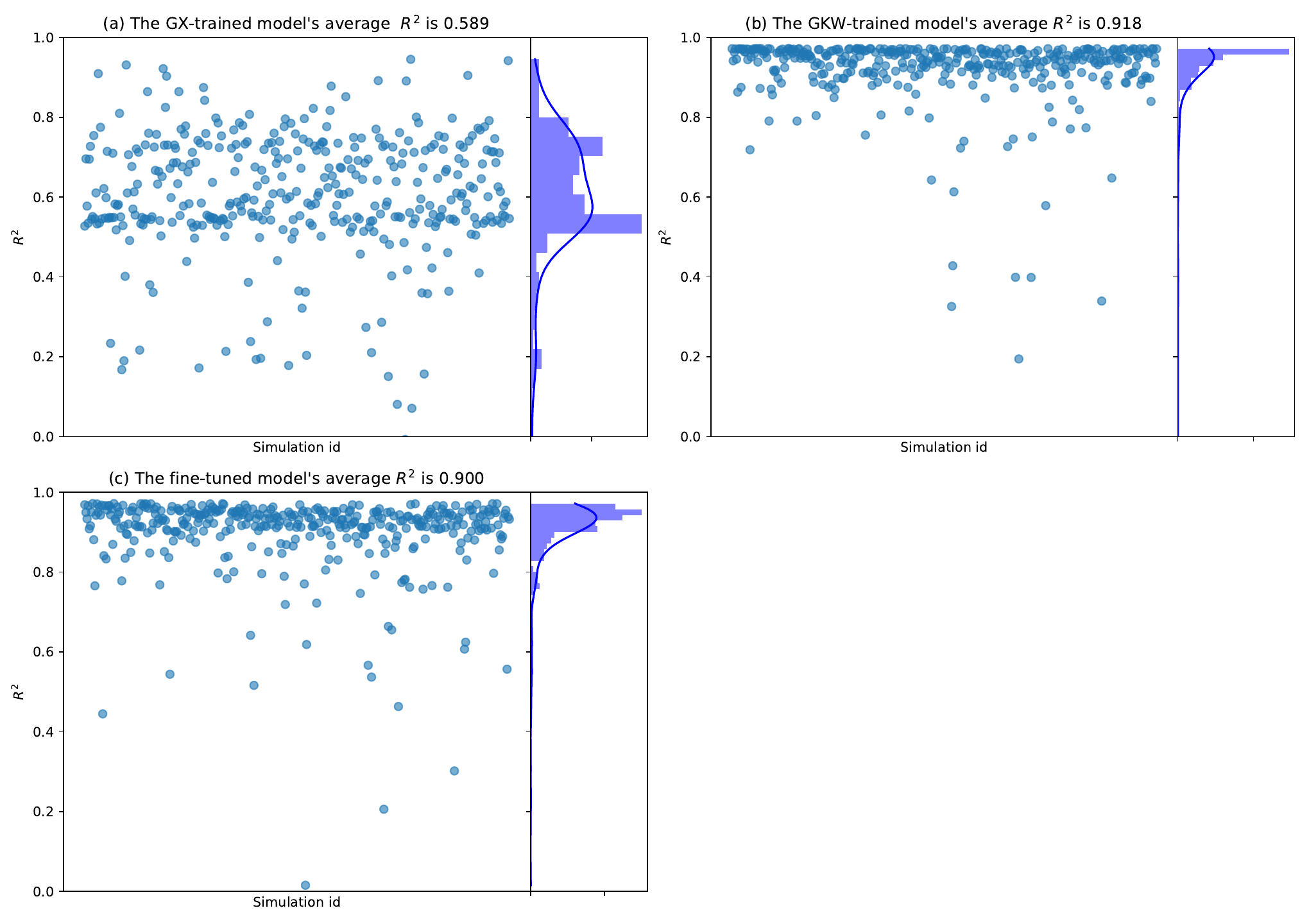}\caption{\label{fig:R2Dis} The $R^{2.pdf}$ distribution in the whole test of
(a) (a) GX-trained, (b) GKW-trained, (c) fine-tuned model predictions.}
\end{figure}

Only one typical result is insufficient to quantitatively evaluate
the accuracy of the model. Therefore, the entire test set $S_{3}$
is used to quantify the accuracy of the surrogate models. Figures
\ref{fig:Regression}, \ref{fig:simDis} and \ref{fig:R2Dis} show
the performance of our models on the entire test set. $R^{2}$ is
coefficient of determination; The definition of similarity is as shown
in equation \ref{eq:1},

\begin{eqnarray}
S\left(x,y\right) & = & \frac{\Sigma(\boldsymbol{x}-\bar{\boldsymbol{x}})(\boldsymbol{y}-\bar{\boldsymbol{y}})}{\sqrt{\Sigma(\boldsymbol{x}-\bar{\boldsymbol{x}})^{2}\Sigma(\boldsymbol{y}-\bar{\boldsymbol{y}})^{2}}}\nonumber \\
\text{similiarity} & = & \frac{S\text{\ensuremath{\left(\hat{\phi_{x}},\phi_{x}\right)}}l_{k_{x}}+S\text{\ensuremath{\left(\hat{\phi_{y}},\phi_{y}\right)}}l_{k_{y}}}{l_{k_{x}}+l_{k_{y}}}\label{eq:1}
\end{eqnarray}

where $\ensuremath{x}$ and $\ensuremath{y}$ are unknown variables,
and $\bar{x}$, $\bar{y}$ are their averages. $l_{k_{x}}$, $l_{k_{y}}$
represent the number of points in $\phi_{x}$ and $\phi_{y}$, respectively,
and $S\left(\boldsymbol{x},\boldsymbol{y}\right)$ denotes the Pearson
correlation coefficient. This equation uses point normalization to
avoid bias in the results due to the differing numbers of points in
$\phi_{x}$ and $\phi_{y}$. $\phi_{x}$ and $\hat{\phi_{x}}$ are
the target and predicted $k_{x}$ spectrum, and the the same notation
applies for $\phi_{y}$ and $\hat{\phi_{y}}$.

From figures \ref{fig:simDis} and \ref{fig:R2Dis}, which illustrate
the similarity distribution on the test set, we observe that the performance
of the model trained with GKW is nearly identical to that of the fine-tuned
model. In contrast, the results of the model trained with GX differ
significantly from the actual GKW simulations. This indicates that
fine-tuning effectively integrates knowledge from both GKW and GX
results. From figure \ref{fig:Regression}, we can observe that the
fine-tuning model prediction is closer to the diagonal than the GX-trained
model prediction, particularly near $\chi_{i}=5$, where the improvement
is most noticeable. However, when $\ensuremath{\chi_{i}}$exceeds
10, the improvement is not satisfactory, likely due to the sparsity
of the fine-tune model's tuning set (20\% $S_{1}$).Actually, the
fine-tuned model training set $20\%\ensuremath{S_{1}}$ is different
from$\ensuremath{S_{3}},$ but they are independent and identically
distributed. \lyxadded{chgwan}{Fri Mar 21 05:38:17 2025}{Figure \ref{fig:Regression}(a)
shows that the predictions of the GX trained model for $\chi_{i}$
exhibit a subset with notably poor performance below the diagonal.
Upon inspection, we find that over 90\% of these points correspond
to cases with $\hat{s}=0.1$. As discussed by \citep{Martin2018,Mandell2022},
the standard twist-shift boundary condition quantizes the perpendicular
aspect ratio as $\ensuremath{L_{x}/L_{y}=J/(2\pi|\hat{s}|)},$ making
resolution particularly demanding for low-shear cases. Although the
generalized twist-shift boundary condition provides some flexibility
in aspect ratio selection, maintaining numerical accuracy at $\ensuremath{\hat{s}\ll1}$remains
challenging. As a result, the GX-trained model exhibits a significant
discrepancy from the target GKW results in the low-$\hat{s}$ regime.
}Also, we examine all similarities lower than 0.8 in figures \ref{fig:simDis}(b),
\ref{fig:simDis}(c), and all$R^{2}$ values lower than 0.8 in \ref{fig:R2Dis}(b)
and \ref{fig:R2Dis}(c). The main cases arise from incorrect simulations,
such as the $\phi_{x}$ spectrum not being strictly symmetrical. The
similarity and $R^{2}$ distributions of the GKW-trained model are
more concentrated than that of the fine-tuned model. We believe that
this is because the fine-tuning used sparser GKW simulation results
than the GKW-trained model. In the future, this could be improved
using more appropriate sampling methods, such as adaptive sampling,
which can sample more points on challenging results.

\section{Discussion and Conclusion \label{sec:Discussion}}

This study demonstrates new possibilities for developing high-fidelity
surrogate models for tokamak simulation codes. In the development
of the ITG surrogate model, by leveraging a large low-fidelity GX
simulation dataset and a small high-fidelity GKW simulation dataset,
we achieve comparable accuracy to the model trained exclusively on
a large high-fidelity GKW simulation dataset while reducing the need
for computationally expensive data by 80\%. The comparison between
fine-tuned model predictions and actual GKW simulations for thermal
diffusivity and wave spectrum was in qualitative agreement. This approach
significantly lowers the barriers to developing surrogate models for
computationally demanding simulations.

Despite these promising results, the study is still a proof-of-principal
work, that warrants further exploration. First, the surrogate model
was trained within a relatively limited parameter space, focusing
on ion temperature gradient (ITG) modes with four input parameters.
Future work should consider expanding this parameter space to include
additional physical phenomena such as trapped electron modes (TEM)
and electron temperature gradient (ETG) modes. Second, it is necessary
to explore more parameters. As noted in this paper, typical gyrokinetic
code simulations require $\sim15$ dimensions of input, and a broader
range of parameters should be scanned in future datasets. Additionally,
while fine-tuning combines datasets of varying fidelity, the results
may be influenced by the specific characteristics of the simulation
codes utilized. Further validation across different simulation frameworks
or experimental datasets would enhance the generalizability of this
method.

The fine-tuning process in this study assumed that low-fidelity GX
simulations capture general patterns, while high-fidelity GKW simulations
refine these patterns. However, the observed overestimation of wave
spectrum in low wavenumber and the underestimation of high $\chi_{i}$
in the fine-tuned surrogate model suggests opportunities to incorporate
adaptive sampling techniques, which dynamically select training points
to cover challenging regions of the parameter space more effectively.
For the thermal diffusivity estimation, we can develop a machine learning\textendash based
classification model to categorize parameters as below or above the
threshold, and then train the regressive surrogate model on parameters
above the threshold. These techniques could further improve model
performance and efficiency.

Finally, the approach outlined here aligns with broader efforts to
develop digital twin systems for tokamaks \citep{Tang2024}. Such
systems require fast, accurate, and high-fidelity simulations to predict
plasma behavior and optimize operational scenarios in real-time. The
fine tuning-based surrogate model provides a foundational framework
for achieving these ambitious goals.

\section*{Data availability statement}

The data that supports the findings of this study belongs to the Singapore
SAFE team and is available from the corresponding author upon reasonable
request.

\ack{}{}

The computational work for this article was partially performed on
resources of the National Supercomputing Centre, Singapore (https://www.nscc.sg).\lyxdeleted{chgwan}{Fri Mar 21 05:55:47 2025}{ }

This research is supported by the National Research Foundation, Singapore,
the China National Postdoctoral Program for Innovative Talents under
Grant No. BX20230371.

\bibliography{library}

\end{document}